\documentclass[12pt]{article}
\input epsf.sty
\usepackage{mathrsfs}
\usepackage{amsmath}
\usepackage{mathrsfs}
\usepackage{amssymb}
\usepackage{color}
\usepackage{psfrag}
\newcommand{\bea}{\begin{eqnarray}}
\newcommand{\eea}{\end{eqnarray}}
\newcommand{\be}{\begin{equation}}
\newcommand{\ee}{\end{equation}}
\newcommand{\vs}[1]{\vspace{#1 mm}}

\newcommand{\dsl}{\pa \kern-0.5em /}

\newcommand{\pa}{\partial}

\newcommand{\nn}{\nonumber\\}

\newcommand{\eqn}[1]{(\ref{#1})}

\begin{document}
\topmargin 0mm
\oddsidemargin 0mm

\begin{flushright}

USTC-ICTS-13-02\\

\end{flushright}

\vspace{2mm}

\begin{center}

{\Large \bf Modulating the phase structure of black D6 branes in
canonical ensemble}

\vs{10}

{\large J. X. Lu\footnote{E-mail: jxlu@ustc.edu.cn} and Ran
Wei\footnote{E-mail: wei88ran@mail.ustc.edu.cn}}

\vspace{4mm}

{\em
 Interdisciplinary Center for Theoretical Study\\
 University of Science and Technology of China, Hefei, Anhui
 230026, China\\}

\end{center}

\vs{10}

\begin{abstract}

There exists a dramatic difference in phase structure between
charged black Dp-branes with $p < 5$ and charged black D5 or D6
branes. In \cite{Lu:2012rm}, we found that the phase structure of
charged black D5 system can be changed qualitatively to that of
charged black Dp with $p < 5$ by adding the delocalized D1 branes
but the similar change does not happen for the charged black D6
system through adding the delocalized D2 branes (Note that both
D1/D5 and D2/D6 belong to the same type of D(p - 4)/Dp system).
Adding the delocalized D4 branes to the black D6 branes doesn't
work, either. In this paper, we consider adding the delocalized D0
branes, the only remaining lower dimensional branes, to the black D6
system for this purpose. We find that the delocalized charged black
D0-branes alone share the same phase structure as the charged black
D6 branes, having no van der Waals-Maxwell liquid-gas type. However,
when the two are combined to form D0/D6, the resulting phase diagram
finally gets changed dramatically to the wanted one, containing now
the above liquid-gas type. This change arises from the interaction
between D6 and the delocalized D0.

\end{abstract}

\newpage
\section{Introduction}
Understanding the nature of black hole thermodynamics may teach us
lessons about quantum gravity. The underlying phase structure, in
particular its possible universality and the associated phase
transitions, can be useful in this regard. For example, if the phase
structure is universal for black holes with different asymptotic
spacetimes (including AdS one), this may hint us that there exits a
general holography for these spacetimes\cite{Carlip:2003ne,
Lundgren:2006kt} (not just true only for AdS case). Motivated by
this, it has been recently shown \cite{Carlip:2003ne,
Lundgren:2006kt} that a large part of the phase structure of a black
hole in asymptotically anti-de Sitter (AdS)
space\cite{Hawking:1982dh, Chamblin:1999tk, Chamblin:1999hg} is not
unique to the AdS black hole, but actually shared universally by
suitably stabilized black holes, say, in asymptotically flat space,
even in the presence of a charge $q$. Concretely, a chargeless
(suitably stabilized) black hole in asymptotically flat space can
also undergo a Hawking-Page transition at certain temperature, now
evaporating into a regular ``hot flat space" instead of a regular
``hot empty AdS space" as for an AdS black hole. Moreover, when $q
\neq 0$, there exists also a critical charge $q_c$ and for $q <
q_c$, the phase diagram universally contains a van der Waals-Maxwell
liquid-gas type phase structure along with a line of first-order
phase transition terminating at a second-order critical point with a
universal critical exponent for specific heat as $2/3$.

An isolated asymptotically flat black hole, unlike the AdS one, is
thermodynamically unstable due to its Hawking radiation and needs to
be stabilized first before its phase structure can be analyzed
properly. The standard practice for this is to place such a system
inside a finite spherical cavity \cite{York:1986it} with its surface
temperature fixed. So a thermodynamical ensemble is considered which
can be either canonical or grand canonical, depending on whether the
charge inside the cavity or the potential at the surface of the
cavity is fixed \cite{Braden:1990hw}. We focus in this paper on the
canonical ensemble, i.e., the charge inside the cavity is fixed.

With the advent of AdS/CFT correspondence, the Hawking-Page
transition for AdS black hole `evaporating' into regular ``hot empty
AdS space" at certain temperature \cite{Hawking:1982dh} corresponds
to the confinement-deconfinement phase transition in large $N$ gauge
theory \cite{Witten:1998zw}. The charged AdS black hole has a phase
structure of van der Waals-Maxwell liquid-gas type, giving rise to
the so-called catastrophic holography as discussed in
\cite{Chamblin:1999tk}. As mentioned above, this kind of phase
structure of AdS black hole is actually universal and is the result
of the boundary condition rather than the exact details of
asymptotical metrics which can be either flat, AdS or dS
\cite{Carlip:2003ne, Lundgren:2006kt}. The boundary condition
realized in each case by the reflecting wall provides actually a
confinement to the underlying system. This may suggest that the AdS
holography is a result of such confinement rather than the detail
properties of AdS space. Then the natural speculation as mentioned
earlier is that a similar holography should hold even in
asymptotically flat space. If such a holography holds indeed, the
natural and interesting questions are: how do we define the
corresponding field theory on the underlying holographic
screen\footnote{For an asymptotically flat black hole without an
origin from branes in string/M theory, establishing such a field
theory description will be extremely difficult, not mentioning the
issue associated with the cavity. For asymptotically flat black
branes, to be discussed next, it is very natural to suppose that
there exists an associated dual field theory arising on the
worldvolume of the corresponding branes. Here the issue is how to
consider properly the cavity effect, which may be viewed as imposing
certain boundary conditions on the fields. Note that such a field
theory, if exists at all, is neither supersymmetric nor conformal in
general, due to the presence of cavity. Pursuing this dual field
theory and the corresponding holography will be our future effort
and what has been discussed for the AdS cases given in
\cite{Chamblin:1999tk, Chamblin:1999hg} should be a good guidance.
Our purpose here and the near-future efforts will be on the
understanding of the potential universal phase structure and its
correlation with the underlying interaction from the gravity side
for branes in string/M theory.} which is supposed to be the
spherical cavity and what do the various thermal dynamical phase
transitions correspond to in the field theory so defined?

We have recently found \cite{Lu:2010xt, Lu:2010au} that the same
characteristic phase structure (including a Hawking-Page transition)
of stabilized uncharged (Schwarzschild) black holes was also shared
by the stabilized uncharged black p-branes in D-dimensional
asymptotically flat spacetime (with the brane worldvolume dimensions
$d = 1 + p\,$) in string/M theory. However, in the charged case, the
van der Waals-Maxwell liquid-gas type phase structure mentioned
earlier, with its characteristic behavior\footnote{The so-called
reduced quantities $b_q (x), \bar b, x, q, q_c$ in Fig. 1 are
defined in \cite{Lu:2010xt} and will also be given later in section
3 for D6-branes. $b_q (x)$ is the reduced inverse of local
temperature of charged black Dp branes under consideration at the
surface of cavity. $\bar b$ is the reduced inverse of the given
temperature of cavity. $x$ is the reduced horizon of the black
system and $q$ is its reduced charge with its corresponding critical
$q_c$. We denote the $\bar x$ as a solution of equation of state
$\bar b = b_q (\bar x)$. In drawing analogy with the usual van der
Waals-Maxwell liquid-gas system, we have here $b_q (x)$ the analog
of pressure, x the analog of volume and q the analog of temperature
in the liquid-gas system. The detail analysis of phase structure and
the related phase transitions for black p-branes is given in
\cite{Lu:2010xt} and will also be briefly discussed later after
\eqn{bbehavior}. In drawing further analogy from the right phase
graph in Fig.1, we have the small stable black p-brane phase the
analog of the liquid phase while the large stable one the analog of
the gas phase in the usual liquid-gas case. } shown in Fig. 1, holds
true only for the stabilized charged black p-branes with the
corresponding $\tilde d \equiv D - d - 2
> 2$ (note that $1 \le \tilde d \le 7$, where $\tilde d + 2$ is the number of the spatial dimensions transverse to the brane).
\begin{figure}
\psfrag{A}{$b_q (x)$} \psfrag{A0}{$b_q (x)$} \psfrag{B}{$x$}
\psfrag{C}{$\bar b$} \psfrag{C0}{$\bar b$} \psfrag{D}{$\bar x$}
\psfrag{E}{$b_{\rm max}$} \psfrag{F}{$b_{\rm min}$} \psfrag{A1}{$q
> q_c$}\psfrag{A2}{$q < q_c$}\psfrag{D1}{$q$}\psfrag{D2}{$1$}\psfrag{C1}{$\bar
x_1$}\psfrag{C2}{$\bar x_2$}\psfrag{C3}{$\bar
x_3$}\psfrag{F1}{$x_{\rm min}$}\psfrag{F2}{$x_{\rm max}$}
\begin{center}
  \includegraphics{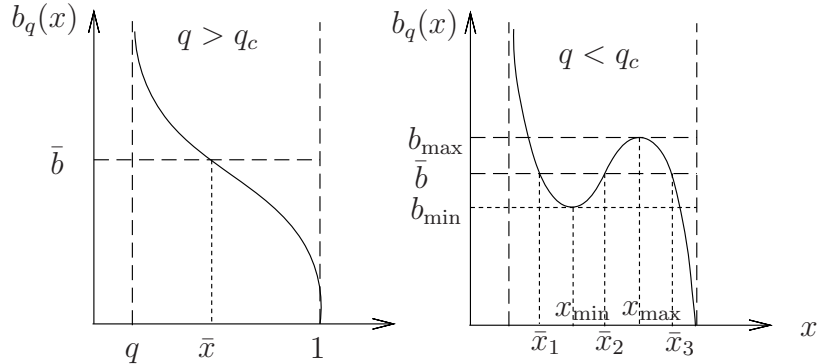}
  \end{center}
  \caption{The typical behavior of $b_q (x)$ vs $x$ for
$q > q_c$ and $q < q_c$}
\end{figure}
 For $\tilde d = 1 ( < 2)$,
the underlying phase structure is qualitative different and actually
resembles that of chargeless case instead. For the $\tilde d = 2$
case, there exists a `critical charge' $q_c = 1/3$ and we have three
subcases to consider, depending on the charge $q > q_c, q = q_c$ and
$q < q_c$ (like the $\tilde d > 2$ cases), but we don't actually
have a critical point in the usual sense and for each subcase the
phase structure looks more like that of the $\tilde d = 1$ case (see
\cite{Lu:2010xt} for detail). Hence this case can be viewed as a
borderline in phase structure which distinguishes the $\tilde d = 1$
case from the $\tilde d > 2$ cases. Since $\tilde d$ is related to
the spatial dimensionality $p$ of the underlying p-branes ($\tilde d
= D - p - 3$), as an example we take $D = 10$ to see their relation.
Then the $\tilde d
> 2$ corresponds to charged black p-branes with $p < 5$
(Dp and/or F-string when p = 1), $\tilde d = 2$ corresponds to
charged black 5-branes (D5 or NS5-branes) and $\tilde d = 1$
corresponds to charged black D6-branes.

Given the above description of phase structure of black p-branes in
string/M theory, we have a puzzle here. When the black p-branes with
different $p$ (with $1 \le \tilde d \le 7$) are uncharged, they all
have the same characteristic phase structure as that of the usual
uncharged (Schwarzschild) black hole, having a Hawking-Page
transition. However, when they are charged, only those black
p-branes with $\tilde d > 2$ share the same characteristic phase
structure of van der Waals-Maxwell liquid-gas type as that of the
usual charged (Reissner-Nordstr${\rm \ddot{o}}$m) black hole. Such a
change of phase structure can be understood as arising from the
repulsive interaction due to the charge added to these branes(in
addition to the already existing gravitational attraction). But then
the natural question would be: why does this kind of change of phase
structure not happen for $\tilde d \le 2$ systems? Understanding
this dramatic difference in phase structure between $\tilde d > 2$
cases and $\tilde d \le 2$ cases can probably teach us lessons not
only about the universality of van der Waals-Maxwell liquid-gas type
phase structure but also the correlation between such universality
and the underlying brane interaction. This may also shed some light
on the possible underlying holography for such stabilized black
hole/brane systems in general if it exists at all.

While a satisfactory understanding or answer of the above needs our
continuing efforts, we expect that a good starting point for this is
to explore the underlying cause why the phase structure of $\tilde d
\le 2$ systems resists the change when the corresponding brane
charges are added\footnote{The obvious thing is that unlike the
$\tilde d
> 2$ cases, the repulsive interaction due to the added charge in the present case
is insufficient to cause the change of the existing phase structure
 and something more is needed.} and whether there
exists means which can be used potentially to change qualitatively
the phase structure of $\tilde d \le 2$ to that of $\tilde d > 2$ in
canonical ensemble.

As a first step, we have been trying to address the second issue,
namely, finding means to change the phase structure of $\tilde d \le
2$ systems to that of the $\tilde d > 2$ ones (when the branes are
charged). Meanwhile we are also accumulating qualitative information
and/or evidence for the first one, namely, the underlying cause for
the resistance of such a change when charges are added, which we
intend  to address quantitatively in detail elsewhere.

We mentioned that the charged black branes with their $\tilde d = 2$
serve as a borderline between the charged black branes with $\tilde
d = 1$ and those with $\tilde d > 2$ in terms of phase structure and
given that fact, it should be much easier to realize such a change
if we focus on the $\tilde d = 2$ branes, which are the charged
black D5 branes (or NS5-branes) in $D = 10$. We found in a recent
publication \cite{Lu:2012rm} that the simplest system to realize
such a change of phase is to add the delocalized charged D-strings
(or F-strings) to the D5 branes (or
NS5-branes)\footnote{\label{footnote1}Adding the delocalized D3
branes will not work and the resulting phase remains the same as
before. This is true also for the general D(p - 2)/Dp system. We
thank Shibaji Roy for participating this unpublished effort. In the
case without the stabilization, this same conclusion was also drawn
earlier in \cite{Cai:2000hn}.}, i.e. to form a D1/D5 (or F/NS5)
system. In addition, we find that the delocalized charged black
D-strings alone have the exact same phase structure as the charged
black D5-branes, without a van der Waals-Maxwell liquid-gas type
structure. This would lead one naively to think that when the two
are combined, the resulting phase structure should remain the same
as before. It turns out that the resulting one is much richer and
contains the liquid-gas type, with now a critical line instead of a
critical point as in the $\tilde d > 2$ cases. We find that the
physical reason for this change is the existence of a non-vanishing
interaction between the delocalized charged black D-strings and
charged black D5-branes, which differs from that between the
delocalized black D-strings or between black D5-branes.

However, in the same study we found that the similar thing does not
happen for the charged black D6 system (corresponding to $\tilde d =
1$) even if we add delocalized charged D2 branes to form D2/D6
system (Note that both D1/D5 and D2/D6 belong to the same type of
D(p - 4)/Dp bound state system). Adding the delocalized D4 branes
 to this system
doesn't work, either (see footnote \ref{footnote1} for this case).
In the present paper, we explore this further by adding the only
remaining lower dimensional branes, namely the delocalized D0
branes, to the charged black D6 system and fortunately, in this case
the required change occurs.

Specifically, we consider adding the charged D0 branes to the D6,
delocalized along  $x^1, x^2, x^3, x^4, x^5, x^6$ of D6 worldvolume
spatial directions. In other words, we consider the system of D0/D6.
In $D = 10$, D0 and D6 are electric-magnetic dual to each other and
unlike the bound state D4/D6 or D2/D6, the extremal configuration of
D0/D6 preserves no supersymmetry and is actually unstable
\cite{Taylor:1997ay, Brandhuber:1997tt, Dhar:1998ip}. But so long
the brane (D0 and D6) charges are large enough\footnote{This should
be true to validate the gravity configuration.}, the corresponding
bound state is long-lived \cite{Dhar:1998ip, Gibbons:1975kk} and
therefore it makes sense to consider its thermodynamical phase
structure.

One may, at first look, attribute the qualitative change of phase
structure of D6 to the added D0 since one would naively think that
for D0 branes, the corresponding $\tilde d = D - 2 - d = 10 - 2 - 1
= 7 > 2$. This is actually not the case. For the delocalized charged
black D0 branes, their own phase structure (without the presence of
D6) is actually the same as that of charged black D6. The same
phenomenon was also noticed for the delocalized charged black
D-strings and the charged black D5 branes considered in
\cite{Lu:2012rm}. This would naively suggest that when both D6 and
the delocalized D0 are present, the resulting phase structure remain
the same as before but this is not the case. Actually, the phase
structure turns out to be much richer, containing the van der
Waals-Maxwell liquid-gas type, dramatically different from the one
when only one type of branes are present. In comparison with the
D1/D5 case, the present phase is even richer and more complicated.
This is due to the complicated parameter relations of the underlying
configuration \cite{Brandhuber:1997tt, Dhar:1998ip, Gibbons:1985ac}.
Our main purpose here is to reveal the dramatic phase change of
charged black D6 when the delocalized charged D0 branes are present
and to gain some qualitative information about the relationship
between the underlying interaction and the corresponding phase
structure. For this, we in this paper consider only a limited case
when the D0 brane charge and D6 brane charge are set to equal and
the dilaton charge vanishes, for which the parameter relations can
be easily solved. We expect to report the full analysis of phase
structure of this system and other related things such as the
quantitative correlation between the underlying interaction and the
phase structure elsewhere.

    This paper is organized as follows. In section 2, we present the
general configuration of charged black D0/D6 given in
\cite{Brandhuber:1997tt} in Euclidean signature, discuss the
thermodynamical stability and obtain the inverse of local
temperature for the purpose of understanding the underlying phase
structure. The details of the underlying phase structure analysis
are discussed in section 3. We explain what causes the dramatic
change of the phase structure of black D6 branes when delocalized
charged D0 are added and conclude this paper in section 4.

\section{The basic setup}
Our focus in this paper is to see how the added delocalized charged
D0 branes (to charged black D6-branes),  in the sense described
above, change the phase structure of the D6-branes qualitatively.
For simplicity, we will not consider here the corresponding bubble
phase as found relevant in \cite{Lu:2011da} in a similar fashion,
which can become globally stable when  certain conditions are met.

Let us consider the configuration\footnote{We actually use the
configuration given in \cite{Brandhuber:1997tt} with some
modifications. The magnetic part of one form $A_1$ has now been
changed to an electric seven-form potential $A_7$ which represents
the presence of D6 branes. In $D = 10$, D0 branes and D6 branes are
electric-magnetic dual to each other.  We also express the metric in
Einstein-frame for our purpose here. We send the charge parameter
$Q, P$ there to $ - Q, - P$ and then without loss of generality
assume $Q > 0, P > 0$ for convenience. We correct a typo in the
electric one-form potential given in \cite{Brandhuber:1997tt} by
replacing the dilaton charge $\Sigma$ with $\Sigma/\sqrt{3}$.} of
charged black D0/D6 \cite{Brandhuber:1997tt, Dhar:1998ip}, now
expressed in Euclidean signature as
 \bea \label{bd0d6} ds^{2}
&=& F A^{-\frac{1}{8}} B^{-\frac{7}{8}} dt^{2}  + \left(B /
A\right)^{\frac{1}{8}}\sum_{i = 1}^{6} dx_i^{2} +
{A}^{\frac{7}{8}}{B}^{\frac{1}{8}} \left( F^{- 1} d \rho^{2} +
\rho^{2} d\Omega_{2}^{2} \right)\nn A_{[1]}&=&i e^{- 3 \phi_{0}/4}\,
Q \left[\frac{1 - \frac{\Sigma}{\rho_+ \sqrt{3}}}{\rho_+ B (\rho_+)}
- \frac{1 - \frac{\Sigma}{\rho \sqrt{3}}}{\rho B (\rho)}\right]dt\nn
A_{[7]}&=&i e^{3 \phi_{0}/4} \, P \left[\frac{1 +
\frac{\Sigma}{\rho_+ \sqrt{3}}}{\rho_+ A (\rho_+)}- \frac{1 +
\frac{\Sigma}{\rho \sqrt{3}}}{\rho A (\rho)}\right]dt\wedge
dx^{1}...\wedge dx^{6}\nn e^{2(\phi-\phi_{0})}&=& \left(B (\rho)/
A(\rho)\right)^{3/2}, \eea where the metric is in Einstein frame,
each form field is obtained following \cite{Braden:1990hw,
Lu:2010xt} in such a way that the form field vanishes at the horizon
so that it is well defined in the local inertial frame, $\phi_0$ is
the asymptotic value of the dilaton, and \bea \label{fab} F (\rho)
&=& \left(1 - \frac{\rho_+}{\rho}\right)\left(1 -
\frac{\rho_-}{\rho}\right),\nn A (\rho) &=& \left(1 -
\frac{\rho_{A+}}{\rho}\right) \left(1 -
\frac{\rho_{A-}}{\rho}\right), \nn B (\rho) &=& \left(1 -
\frac{\rho_{B+}}{\rho}\right) \left(1 -
\frac{\rho_{B-}}{\rho}\right),\eea with \bea \label{rho-pm} \rho_\pm
&=& M \pm \sqrt{M^2 + \Sigma^2 - P^2/4 - Q^2/4},\nn \rho_{A\pm} &=&
\frac{\Sigma}{\sqrt{3}} \pm \sqrt{\frac{P^2 \Sigma/2}{\Sigma -
\sqrt{3} M}},\nn \rho_{B\pm} &=& - \frac{\Sigma}{\sqrt{3}} \pm
\sqrt{\frac{Q^2 \Sigma/2}{\Sigma + \sqrt{3} M}}.\eea In the above,
the solutions are characterized by the mass parameter $M$, the
delocalized D0 charge parameter $Q$ and the D6 charge parameter $P$.
The dilaton charge parameter $\Sigma$ is not independent and is
related to $M, Q, P$ via the following \be \label{dilatoncharge}
\frac{8}{3}\Sigma = \frac{Q^2}{\Sigma + \sqrt{3} M} +
\frac{P^2}{\Sigma - \sqrt{3} M}.\ee As noticed in
\cite{Brandhuber:1997tt}, under the electric-magnetic duality, the
parameters for the solutions are transformed in the following way
\be \label{emd} Q \leftrightarrow P,\quad \Sigma \leftrightarrow -
\Sigma,\quad M \leftrightarrow M.\ee While the above general
configuration is formally given in terms of three parameters $M, P,
Q$, it doesn't mean that each set of given values of them, even
satisfying \eqn{dilatoncharge}, will give a well-defined
configuration suitable for thermodynamical purpose. For example, it
could give a configuration with a naked singularity for certain
choice of parameters\footnote{For certain values of $M, Q, P$,
either $\rho_{A+}$ or $\rho_{B+}$ can be greater than $\rho_+$, then
the corresponding configuration will have a naked singularity. For
example, for the case of $Q = P$, in addition to the $\Sigma = 0$
solution considered in this paper, we have the other two solutions
for $\Sigma$ from \eqn{dilatoncharge}, either of which gives a
configuration with a naked singularity, therefore  not suitable for
 thermodynamics.}. For the purpose of the present paper mentioned
in the Introduction, we will limit to consider three special cases,
namely, 1) the charged black D6-branes, 2) the delocalized charged
black D0-branes and 3) the special case of $P = Q$ with vanishing
dilaton charge ($\Sigma = 0$), each of the configurations is
well-defined and has a clear meaning of horizon at $\rho = \rho_+$.
We hope to return to give a full analysis of the underlying phase
structure elsewhere for the remaining well-defined configurations
with allowed parameters $M, P, Q$ and the suitable solution for
dilaton charge $\Sigma$ from \eqn{dilatoncharge}. For the three
special cases just mentioned, each of them in Lorentzian signature
has a well-defined horizon at radial coordinate $\rho = \rho_+$. Let
us consider them each in order:
\subsection{Case 1: Q= 0}
Let us first bring the configuration given by
\eqn{bd0d6}-\eqn{dilatoncharge} in this case to the familiar charged
black D6 one \cite{Horowitz:1991cd, Duff:1993ye}. When $Q = 0$, we
have from \eqn{fab} as \bea \label{fabzeroq}F (\rho) &=& \left(1 -
\frac{\rho_+}{\rho}\right)\left(1 - \frac{\rho_-}{\rho}\right),\nn A
(\rho) &=& \left(1 + \frac{\Sigma}{\sqrt{3}\, \rho}\right) \left(1 -
\frac{\sqrt{3}\Sigma}{\rho}\right), \nn B (\rho) &=& \left(1 +
\frac{\Sigma}{\sqrt{3}\,\rho}\right)^2,\eea where we have used
\eqn{dilatoncharge} and \eqn{rho-pm}. Before we give the explicit
$\rho_\pm$, let us examine the expression for the dilaton given in
\eqn{bd0d6} for the present case and we know in a proper radial
coordinate $r$ \cite{Horowitz:1991cd, Duff:1993ye}\be e^{4 (\phi -
\phi_0)/3} = \triangle_- (r) \equiv 1 - \frac{r_-}{r}.\ee We then
have the identifications \be \label{r-rho-t} r = \rho - \sqrt{3}\,
\Sigma, \quad r_- = - \frac{4\Sigma}{\sqrt{3}} > 0, \ee with a
negative dilaton charge as a solution from \eqn{dilatoncharge}. It
is more convenient using parameters $\Sigma, P$ instead of $M, P$.
Then we have from \eqn{dilatoncharge} \be M =
\frac{\Sigma}{\sqrt{3}} - \frac{\sqrt{3}}{8} \frac{P^2}{\Sigma},\ee
which is supposed to be positive. With this, we have
consistently\footnote{Consistency requires $0 > \Sigma > -
\frac{\sqrt{3}}{4} P$ with $P > 0$.} $\rho_\pm$ from \eqn{rho-pm} as
\be \rho_+ = \sqrt{3}\, \Sigma - \frac{\sqrt{3}}{4}
\frac{P^2}{\Sigma}, \quad \rho_- = - \frac{\Sigma}{\sqrt{3}}.\ee We
now use the first equation for the coordinate relation given in
\eqn{r-rho-t} to have $r_\pm = \rho_\pm - \sqrt{3}\, \Sigma$ as \be
\label{r-pm} r_+ = - \frac{\sqrt{3}}{4} \frac{P^2}{\Sigma} > 0,
\quad r_- = - \frac{4\Sigma}{\sqrt{3}}
> 0,\ee where the expression for $r_-$ is just the one given early in \eqn{r-rho-t},
therefore a consistent check. Note also that \be\label{r-pm-p} r_+
r_- = P^2,\ee an expected result. In terms of the radial r
coordinate, we have from \eqn{fabzeroq} \bea \label{fabzeroq}F (r)
&=& \triangle_+ \triangle_- \left(1 + \frac{\sqrt{3}
\Sigma}{r}\right)^{-2},\nn A (r) &=& \triangle_- \left(1 +
\frac{\sqrt{3} \Sigma}{r}\right)^{-2}, \nn B (r) &=& \triangle_-^2
\left(1 + \frac{\sqrt{3} \Sigma}{r}\right)^{-2},\eea where \be
\label{delta-pm} \triangle_\pm = 1 - \frac{r_\pm}{r},\ee with
$r_\pm$ given in \eqn{r-pm}. We then have the configuration
\eqn{bd0d6} in the present case as \bea \label{bd6} ds^{2} &=&
\triangle^{1/8}_-\left( \frac{\triangle_+}{\triangle_-} dt^{2}  +
\sum_{i = 1}^{6} dx_i^{2}\right) + \triangle^{- 7/8}_-
\left(\frac{\triangle_-}{\triangle_+} d r^{2} + r^{2} \triangle^2_-
d\Omega_{2}^{2} \right)\nn A_{[1]}&=& 0\nn A_{[7]}&=&i e^{3
\phi_{0}/4} \, P \left[\frac{1}{r_+} - \frac{1}{r}\right]dt\wedge
dx^{1}...\wedge dx^{6}\nn e^{2(\phi-\phi_{0})}&=& \triangle_-^{3/2},
\eea which is just the standard black D6 brane configuration now
expressed in Euclidean signature, for example, see
\cite{Duff:1994an}, with $P = (r_+ r_-)^{1/2}$.
\subsection{Case
2: P = 0} We also want to bring the delocalized charged black D0
brane configuration to the form similar to that given in the
previous subsection for the charged black D6 brane one. Given the
electric-magnetic relation \eqn{emd}, we expect that for this case
the dilaton charge should be positive which is true indeed. For this
case, we have from \eqn{fab} as \bea \label{fabzerop} F (\rho) &=&
\left(1 - \frac{\rho_+}{\rho}\right)\left(1 -
\frac{\rho_-}{\rho}\right),\nn A (\rho) &=& \left(1 -
\frac{\Sigma}{\sqrt{3}\,\rho}\right)^2, \nn B (\rho) &=& \left(1 -
\frac{\Sigma}{\sqrt{3}\,\rho}\right) \left(1 +
\frac{\sqrt{3}\,\Sigma}{\rho}\right),\eea where we also use
\eqn{dilatoncharge} and \eqn{rho-pm} for the present case. Once
again before we give explicit expressions for $\rho_\pm$, we would
like to express the configuration in a proper radial coordinate $r$
such that \be e^{4 (\phi - \phi_0)/3} = \triangle^{-1}_- \equiv
\left(1 - \frac{r_-}{r}\right)^{-1}.\ee Using the dilaton expression
in \eqn{bd0d6} and the functions $A (\rho)$ and $B (\rho)$ given in
\eqn{fabzerop}, we can read \be \label{r-rho-tq} r = \rho +
\sqrt{3}\, \Sigma, \quad r_- = \frac{4 \Sigma}{\sqrt{3}} > 0,\ee
with now $\Sigma > 0$. One can also see that these two relations can
be obtained from \eqn{r-rho-t} by sending $\Sigma \rightarrow -
\Sigma$. Once again it is more convenient using parameter $\Sigma,
Q$ than $M, Q$. We can solve $M$ in terms of $\Sigma, Q$ from
\eqn{dilatoncharge} as \be M = - \frac{\Sigma}{\sqrt{3}} +
\frac{\sqrt{3}}{8} \frac{Q^2}{\Sigma} > 0.\ee With this, we can have
consistently\footnote{Consistency requires now $0 < \Sigma <
\sqrt{3} Q /4$ with $Q > 0$.} from \eqn{rho-pm} \be \rho_+ =
\frac{\sqrt{3}}{4} \frac{Q^2}{\Sigma} - \sqrt{3}\, \Sigma > 0, \quad
\rho_- = \frac{\Sigma}{\sqrt{3}} > 0.\ee From the first expression
given in \eqn{r-rho-tq}, we have then \be \label{rpmq}r_+ =
\frac{\sqrt{3}}{4} \frac{Q^2}{\Sigma} > 0, \quad r_- =
\frac{4\Sigma}{\sqrt{3}} > 0,\ee where $r_-$ is just the one given
earlier in \eqn{r-rho-tq}, a consistent check. Note also now that
\be \label{r-pm-q} r_+ r_- = Q^2,\ee an expected result. In terms of
the radial coordinate $r$, we have from
\eqn{fabzerop}\bea\label{fabrzerop} F (r) &=& \triangle_+
\triangle_- \left(1 - \frac{\sqrt{3} \, \Sigma}{r}\right)^{- 2},\nn
A (r) &=& \triangle_-^2 \left(1 - \frac{\sqrt{3} \,
\Sigma}{r}\right)^{- 2} \nn B (r) &=& \triangle_- \left(1 -
\frac{\sqrt{3} \, \Sigma}{r}\right)^{- 2},\eea where $\triangle_\pm$
are given by \eqn{delta-pm} but now with $r_\pm$ given in
\eqn{rpmq}. Then the delocalized charged black D0 brane
configuration from \eqn{bd0d6} is \bea \label{bd0} ds^{2} &=&
\triangle_-^{- 1/8}\left(\triangle_+  dt^{2} + \sum_{i = 1}^{6}
dx_i^{2}\right) + \triangle_-^{7/8} \left( \frac{ d
r^{2}}{\triangle_+} + r^{2} \triangle_- d\Omega_{2}^{2} \right)\nn
A_{[1]}&=&i e^{- 3 \phi_{0}/4}\, Q \left[\frac{1}{r_+} - \frac{1}{r}
\right]dt\nn A_{[7]}&=& 0, \nn e^{2(\phi-\phi_{0})}&=&
\triangle_-^{- 3/2}.\eea This explicit expression, to our knowledge,
has not been written down before, which is related to \eqn{bd6} by
T-dualities along the isometric directions $x^i$ ($i = 1, 2,\cdots
6$) with the parameter $P$ replaced by the present $Q$ (Note that
the delocalized D0 configuration \eqn{bd0} has the same isometries
as the charged black D6 configuration \eqn{bd6}).
\subsection{Case 3: P = Q, $\Sigma = 0$}
This is the case which we will use to demonstrate how the added
delocalized charged D0 branes change the phase structure of charged
black D6 branes with $\tilde d = 1$ dramatically to the one similar
to that of charged black p-branes with $\tilde d
> 2$ in canonical ensemble discussed in \cite{Lu:2010xt}, containing a
van-der Waals liquid-gas phase structure. For $P = Q = K$, one of
solutions from \eqn{dilatoncharge} is $\Sigma = 0$, whose extremal
one is given in \cite{Sheinblatt:1997nt}. We have now from \eqn{fab}
\be \label{fabpq} F (r) = \triangle_- \triangle_+,\quad A (r) = 1,
\quad B (r) = 1,\ee where we have set $r = \rho$ for the present
case and $\triangle_\pm$ are once again given in \eqn{delta-pm} but
with now $r_\pm$ given as \be r_\pm = M \pm \sqrt{M^2 - K^2/2},\ee
which gives \be \label{r-pm-pq} r_+ r_- = K^2/2.\ee We have now the
configuration from \eqn{bd0d6} as \bea \label{bd0d6k} ds^{2} &=&
\triangle_- \triangle_+ dt^{2} + \sum_{i = 1}^{6} dx_i^{2} + \frac{d
r^2}{\triangle_- \triangle_+} + r^{2} d\Omega_{2}^{2},\nn A_{[1]}&=&
i e^{- 3 \phi_{0}/4}\, K \left[\frac{1}{r_+} - \frac{1}{r}\right]
dt\nn A_{[7]}&=& i e^{3 \phi_{0}/4} \, K \left[\frac{1}{r_+} -
\frac{1}{r} \right]dt\wedge dx^{1}...\wedge dx^{6}\nn
e^{2(\phi-\phi_{0})}&=& 1. \eea For this configuration, the dilaton
remains as a constant and it looks like the usual 4-dimensional
Reissner-Nordstr\"{o}m solution, having two horizons, the outer one
at $r = r_+$ and inner one at $r = r_-$, with a singularity at $r =
0$. For the purpose of this paper on phase structure, we are
interested only in the outer horizon.

We now return to the general configuration \eqn{bd0d6} and assume
from now on that this configuration  is well-defined with a horizon
(in Lorentizian signature) at $\rho = \rho_+$. For this metric to be
free from the conical singularity at the horizon, the Euclidean time
coordinate `t' must be periodic with periodicity \be \beta^* =
\frac{4 \pi \rho^2_+ \sqrt{A (\rho_+) B (\rho_+)}}{\rho_+ - \rho_-},
\ee the inverse of temperature of the black D0/D6 system at $\rho =
\infty$. Here $A(\rho_+), B(\rho_+)$ are the ones given in \eqn{fab}
with $\rho = \rho_+$.  With this,  we can read, from the metric, the
inverse of local temperature at a given $\rho$ as \be \label{localt}
\beta = F^{1/2} A^{- 1/16} B^{- 7/16} \beta^*.\ee Applying to the
three specific cases considered, we have the inverse of local
temperature at a given $r$ for Case 1, \be \label{case1beta}\beta =
4 \pi r_+ \triangle_+^{1/2} (r) \triangle_-^{- 7/16} (r) \left(1 -
\frac{r_-}{r_+}\right)^{1/2} = 4 \pi \bar r_+ \triangle_-^{-1} (\bar
r) \triangle_+^{1/2} (\bar r) \left(1 - \frac{\bar r_-}{\bar
r_+}\right)^{1/2},\ee for Case 2, \be\label{case2beta} \beta = 4 \pi
r_+ \triangle_-^{- 1/16} (r) \triangle_+^{1/2} (r) \left(1 -
\frac{r_-}{r_+}\right)^{1/2} = 4\pi \bar r_+ \triangle_-^{-1} (\bar
r) \triangle_+^{1/2} (\bar r) \left(1 - \frac{\bar r_-}{\bar
r_+}\right)^{1/2},\ee and for case 3, \be \label{case3beta} \beta =
4\pi r_+ \triangle_-^{1/2} (r) \triangle_+^{1/2} (r) \left(1 -
\frac{r_-}{r_+}\right)^{-1} = 4\pi \bar r_+ \triangle_-^{1/2} (\bar
r) \triangle_+^{1/2} (\bar r) \left(1 - \frac{\bar r_-}{\bar
r_+}\right)^{-1},\ee where for each case in the last equality, we
have used the so-called physical radius $\bar r$ (rather than the
coordinate one $r$) via $\bar r = r \triangle_-^{ 9/16}$ for Case 1,
$\bar r = r \triangle_-^{15/16}$ for Case 2, and $\bar r = r$ for
Case 3, respectively, from their corresponding metric. In the above,
we also define the physical parameters $\bar r_\pm$ in the same way
for each case and the corresponding $\triangle_\pm (\bar r)$ remains
in form the same as before, \be \triangle_\pm = 1 - \frac{r_\pm}{r}
= 1 - \frac{\bar r_\pm}{\bar r}.\ee

To study the equilibrium thermodynamics \cite{Gibbons:1976ue} in
canonical ensemble, the allowed configuration (say, the black D0/D6
system or its extremal one) should be placed in a cavity with a
fixed radius $\bar r_B$ ($> \bar r_+$) for the reason as explained
in the Introduction \cite{York:1986it}. The other fixed quantities
are the cavity temperature $1/\beta$, the physical periodicity of
each $x^i$ with\footnote{Note that to have a finite Euclidean action
for the D0/D6 system, the brane coordinates $x^i$ with $i = 1,
\cdots 6$ should be compact.} $i = 1, \cdots 6$, the dilaton value
$\bar \phi$ on the surface of the cavity (at $\bar r = \bar r_B$)
and the charges/fluxes enclosed in the cavity $\bar Q_p$ ($p = 0,
6$), respectively. In equilibrium, these fixed values are set equal
to the corresponding ones of the allowed configuration enclosed in
the cavity. For example, we set the charge \be\label{charge} \bar
Q_p = Q_p \equiv \frac{i}{\sqrt{2} \kappa}\int e^{-a(d)\phi} \ast
F_{[d + 1]},\ee for $d = p + 1 = 1, 7$, respectively. In the above,
$\ast$ denotes the Hodge duality and the field strength $F_{[d +
1]}=  d A_{[d]}$ with $A_{[d]}$ the corresponding form potential.
With the potentials $A_1, A_7$ given in \eqn{bd0d6}, we have \bea
F_2 &=&  i
e^{- 3 \phi_0/4} Q \frac{A (\rho)}{\rho^2 B^2 (\rho)} d \rho\wedge dt, \\
 F_8 &=&  i e^{3\phi_0/4} P \frac{B (\rho)}{\rho^2 A^2 (\rho)} \,
 d\rho
 \wedge dt \wedge d x^1 \wedge \cdots \wedge d x^6,\eea where the explicit expressions for $A
 (\rho)$ and $B (\rho)$ are given in \eqn{fab}, respectively.
 We therefore have the D0-brane charge and D6-brane
charge per unit six-brane volume, respectively, as
 \be Q_0 = \frac{\Omega_2 Q}{\sqrt{2} \kappa} e^{3
 \phi_0/4} V_6^*, \quad Q_6 = \frac{\Omega_2 P}{\sqrt{2} \kappa}
 e^{- 3\phi_0/4},\ee
 where $\Omega_n$ denotes the volume of a unit $n$-sphere, $V_6^* \equiv \int d x^1 d x^2 \cdots d x^6$ is the coordinate volume, and $\kappa$
is a constant with $1/ (2 \kappa^2)$ appearing in front of the
Hilbert-Einstein action in canonical frame but containing no
asymptotic string coupling $g_s = e^{\phi_0}$. For the consideration
of phase structure, as usual, we need to express these charges in
each case in terms of their fixed or/and physical quantities on the
cavity. We have for Case 1, \be \label{case1charge} Q_0 = 0, \quad
Q_6 = \frac{\Omega_2}{\sqrt{2}\kappa} e^{- 3\bar\phi/4} (\bar r_+
\bar r_-)^{1/2},\ee where we have used the dilaton formula in
\eqn{bd6} to express its asymptotic value $\phi_0$ in terms of its
fixed value $\bar \phi$ at $\bar r = \bar r_B$ and use the physical
parameter $\bar r_\pm$ in replace of $P$ via \eqn{r-pm-p}, for Case
2, \be \label{case2charge} Q_0 = \frac{\Omega_2 \bar
V_6}{\sqrt{2}\kappa} e^{3\bar\phi/4} (\bar r_- \bar r_+)^{1/2},
\quad Q_6 = 0,\ee where we have used  $\bar\phi$ in replace of
 $\phi_0$ via \eqn{bd0} and the physical parameter $\bar r_\pm$ in replace of $Q$ via \eqn{r-pm-q}, respectively,
   and the physical volume $\bar V_6 = \triangle_-^{- 3/8} V_6^*$,
and for Case 3, \be \label{case3charge}Q_0 = \frac{\Omega_2 \bar
V_6}{\sqrt{2} \kappa} e^{3 \bar\phi/4} (2 \bar r_-\bar r_+)^{1/2},
\quad Q_6 = \frac{\Omega_2}{\sqrt{2} \kappa}
 e^{- 3\bar \phi/4} (2\bar r_-\bar r_+)^{1/2},\ee where now $\bar
 \phi = \phi_0$, $\bar V_6 = V_6^*$ and we have expressed $P = Q =
 K$ in terms of $\bar r_\pm$ using \eqn{r-pm-pq} (note here $\bar
 r_\pm = r_\pm$).

 In the above, we
have expressed all the quantities in terms of their fixed (or
physical) correspondences, for example, we have replaced the
asymptotical string coupling $g_s$ in terms of the fixed effective
string coupling on the cavity via $e^{\bar\phi} = e^{\phi(\bar
r_B)}$ in each case.  In canonical ensemble, it is the Helmholtz
free energy which determines the stability of equilibrium states and
is related to the Euclidean action by $F=I_E/\beta$ in the leading
order approximation. The procedure for evaluating the Euclidean
action of black $p$-branes was given in detail in \cite{Lu:2010xt}
following the standard technique. The generalization to the present
case is straightforward, similar to what  we did for the D(p - 4)/Dp
systems in \cite{Lu:2012rm}, though the computation is a bit
lengthy, by considering one more piece contribution from the form
field strength $F_2$ and its potential $A_1$ in addition to that
from the usual $F_8$ and its potential $A_7$ for black D6. The
explicit expression of action for each relevant case will be given
in the following section and it can be used to analyze the
underlying phase structure.  What is important is the recognition
that for given cavity temperature $1/\beta$, the Euclidean action is
essentially the free energy ($I_E = \beta F$). So the minimum of
free energy is that of the Euclidean action. Hence the minimum of
Euclidean action will determine the local stability of the
underlying system under consideration. Note that in canonical
ensemble, the only variable is the horizon size $\bar r_+$ and we
always have\footnote{Another way to understand this is $I_E (r_+) =
\beta F = \beta E (r_+) - S (r_+)$ where $E (r_+)$ is the internal
energy and $S(r_+)$ the entropy. In canonical ensemble, we have
fixed cavity size, temperature and the charge inside, so $d E = T
(r_+) d S$. We then have $d I_E/d r_+ = \beta d E/d r_+ - d S/d r_+
= T (r_+) (d S/d r_+) (\beta - \beta (r_+))$ with $\beta (r_+) =
1/T(r_+)$. Note that $S$ is the only function of $r_+$ and $d S/d
r_+ > 0$, therefore we have \eqn{stability}.} \be \label{stability}
\frac{d I_E (\bar r_+)}{d \bar r_+} \sim (\beta - \beta (\bar
r_+)),\ee and \be \label{eos} \frac{d I_E (\bar r_+)}{d \bar r_+} =
0 \Rightarrow \beta (\bar r_+) = \beta,\ee where the extremal
condition of $I_E$ is nothing but the thermal equilibrium of the
system inside cavity with the cavity with a preset temperature
$1/\beta$. In the above, we have computed the respective $\beta
(\bar r_+)$ using the corresponding explicit Euclidean action given
in the following section and find that it is nothing but the inverse
of local temperature at $\bar r = \bar r_B$ for each case given
earlier in \eqn{case1beta}, \eqn{case2beta} and \eqn{case3beta},
respectively, therefore a consistent check. At the solution of
$\beta (\bar r_+) = \beta$ given above, we have \be \frac{d^2 I_E
(\bar r_+)}{d \bar r_+^2} \sim  - \frac{d \beta (\bar r_+)}{d\bar
r_+},\ee so the minimum of $I_E$ at the solution, i.e., $d^2 I_E/d
\bar r_+^2 > 0$, requires the negative slope of $\beta (\bar r_+)$
there. Therefore, as always, the function $\beta (\bar r_+)$
computed above or given earlier is the key for us to determine the
phase structure of underlying system which we will discuss in the
following section. The above stability discussion can be verified
for each case considered with the corresponding explicitly-evaluated
Euclidean action given in the following section and the key function
$\beta (\bar r_+)$ for phase structure can also be obtained
accordingly.

\section{The analysis of phase structure}

In this section, we will give the phase structure of each case
considered in the previous sections, in particular, Case 3 which is
the focus of this paper. Let us now discuss each of them in order.
\subsection{Case 1: Q = 0}
This is just the usual charged black D6 brane and the inverse of
local temperature given in \eqn{case1beta} is nothing but the one
given in \cite{Lu:2010xt}. The phase structure of this system, i.e,
the $\tilde d = 1$ case, was given there, similar to the one for
chargeless black p-branes, containing no the van-der Waals-Maxwell
liquid-gas type. We will refer this case to \cite{Lu:2010xt} for
detail.

\subsection{Case 2: P = 0}

   This is the first new case encountered in this paper.  The delocalized
charged black D0 branes alone would at first look give a rather
different phase structure than the charged black D6 branes since the
localized charged black D0 branes seemly correspond to the $\tilde d
= 7$ ($\tilde d = D - d - 2 = 10 - 1 - 2 = 7$) case discussed in
\cite{Lu:2010xt},  containing the van der Waals-Maxwell liquid-gas
phase structure. However, just like the delocalized D-strings in the
case of D1/D5 discussed very recently in \cite{Lu:2012rm}, the
present delocalized black D0 branes once again turn out to share the
same phase structure as the charged black D6 branes, containing
actually no van der Waals-Maxwell liquid-gas type. This can be seen
from the inverse of local temperature in canonical ensemble for the
present case given in \eqn{case2beta} which is identical to the one
for the charged black D6 branes as given in \eqn{case1beta}. This
can also be seen from its explicit reduced Euclidean action \bea
\tilde I_E &\equiv& \frac{2 \kappa^2 I_E}{4 \pi \Omega_2 \bar V_6
\bar r_B^2}\nn &=& \frac{\beta}{4\pi \bar r_B}\left[ - 3
\sqrt{\frac{\triangle_+}{\triangle_-}} - \sqrt{\triangle_+
\triangle_-} + 4\right] - \left(\frac{\bar r_+}{\bar r_B}\right)^2
\triangle_-^{- 3/2} \left(1 - \frac{\bar r_-}{\bar
r_+}\right)^{3/2},\eea which is identical in form to the
correspondence of charged black D6 branes given in \cite{Lu:2010xt}
and from which the same $\beta (\bar r_+)$ as given in
\eqn{case2beta} can be obtained following the prescription given in
the previous section. This is a bit surprise since the metric as
given in \eqn{bd0} for the delocalized charged black D0 branes looks
completely different from that given in \eqn{bd6} for charged black
D6 branes, not mentioning two types of different objects. One of
possible reasons for the two sharing the same phase structure is
that from our experience on the black p-branes given in
\cite{Lu:2010xt}, the underlying phase structure depends crucially
on the dimensionality of transverse dimensions and for both of these
cases this dimensionality is indeed the same. The other more
sounding reason is that these two configurations are related to each
other by T-dualities\footnote{Showing this is straightforward,
following\cite{Bergshoeff:1995as,Breckenridge:1996tt,Lu:2005ju}. We
stress that an isometric direction transverse to a black brane
cannot be generated by placing an infinite periodic array of such
black branes along this direction since unlike the BPS case we do
not have the so-called ``no-force" condition, crucial for this
purpose, for black branes placed parallel to each other. In general,
one has to solve the corresponding equations of motion for such
delocalized black brane configuration with certain isometries for
performing T-dualities along these directions, see \cite{Lu:2005ju}
for example. Two black configurations can be T-dual related to each
other only when the two configurations have the same isometries.}
along the isometric $x^i$ directions with $i = 1, \cdots 6$,
therefore are T-dual equivalent(Note that the delocalized black D0
brane configuration \eqn{bd0} has the same isometries as the charged
black D6 brane configuration \eqn{bd6}) and T-duality keeps the
underlying phase structure unchange.
\subsection{Case 3: P = Q = K}

  This case is the main focus of this paper reflecting the dramatic change
  of phase structure of the charged black D6 branes when the
  delocalized charged D0 branes are added. The discussion of the previous two
  subsections seemly suggests that when the delocalized charged D0
  branes and charged D6 branes are combined to form the charged black
  D0/D6 system, the resulting phase structure would be
  supposed to remain the same as that when either type of the branes are
  present. This actually happens when the delocalized D4 branes
  are added to the charged black D6 branes as mentioned in footnote
  \ref{footnote1} and this remains almost true when the delocalized
  D2 branes are added\footnote{When the delocalized charged D2 branes are added to the charged black D6 branes
  \cite{Lu:2012rm}, the resulting phase has been changed to one similar to
  that of the charged black D5 even though the basic phase structure
  is still the one similar to the chargeless case.}. The only remaining lower
  dimensional objects available for our purpose are the delocalized
  D0 branes. Just like the added delocalized D-strings to the
  charged  black D5 branes, the added delocalized D0-branes finally
  change the phase structure of charged black D6 branes to the one
  similar to that of black p-branes with $\tilde d > 2$, containing a van
  der Waals-Maxwell liquid-gas type. Let us examine this in what
  follows.

  The key quantity as mentioned at the end of last section is the
  inverse of local temperature $\beta (r_+)$ given in
  \eqn{case3beta}. Before discussing the behavior of this function,
  we address a few related issues. This case is $P = Q = K$ and from
  \eqn{case3charge} and \eqn{r-pm-pq} we need to set \be \label{charge-relation} \frac{\bar Q_0
  }{\bar V_6} =
  \bar Q_6 \, e^{3\bar \phi/2},\ee where we have set $Q_p = \bar Q_p$ for $p = 0, 6$, respectively.
   Note that $\bar Q_0, \bar Q_6,
  \bar\phi$ and $\bar V_6$ are all fixed in the canonical ensemble
  and for the present case they need to satisfy the above relation.
  Using \eqn{case3charge}, we define a reduced fixed charge with a
  dimension of length as
  \be\label{reducedcharge} \tilde Q  = \frac{\kappa\, \bar Q_6\,
  e^{3\bar \phi/4}}{\Omega_2} = \frac{\kappa\, \bar Q_0\, e^{ -
  3\bar\phi/4}}{\bar V_6\, \Omega_2}.\ee With this, we have from
  \eqn{case3charge}
  \be\label{r-m}  \bar r_- = \frac{\tilde Q^2}{\bar r_+}, \ee which
  indicates that $\bar r_-$ is not independent and is related to
  $\bar r_+$ via the above. For simplicity of further discussion, as usual, we
  define the so-called reduced quantities at the fixed radius $\bar r = \bar r_B$,
  \be \label{reducedquantities} x \equiv \frac{\bar r_+}{\bar r_B} < 1, \quad \bar b \equiv
  \frac{\beta}{4\pi \bar r_B}, \quad q \equiv \frac{\tilde Q}{\bar
  r_B} < x.\ee Note that $0 < q < 1$ (since $x < 1$) and $q < x < 1$ (since $\bar r_+ > \bar r_-$). With these, we have
  \be \label{deltas} \triangle_- = 1 - \frac{\bar r_-}{\bar r_B} = 1 -
  \frac{q^2}{x},\quad \triangle_+ = 1 - \frac{\bar r_+}{\bar r_B} =
  1 - x.\ee

  At $\bar r = \bar r_B$, the inverse of local temperature \eqn{case3beta} for the
  present case is
  \be \label{case3beta-cavity} \beta (\bar r_+) =
  4\pi \bar r_+ \triangle_-^{1/2} (\bar r_B) \triangle_+^{1/2} (\bar r_B)
\left(1 - \frac{\bar r_-}{\bar r_+}\right)^{-1}.\ee Therefore, from
  the above, the corresponding inverse of reduced
  local temperature is
  \be \label{reducedb} b_q (x) \equiv \frac{\beta (r_+)}{4\pi r_B} =
  \frac{x (1 - x)^{1/2} \left(1 - \frac{q^2}{x}\right)^{1/2}}{1 -
  \frac{q^2}{x^2}}.\ee
  This same $b_q (x)$ can also be obtained from the corresponding
  reduced Euclidean action given below, following the prescription given in the
  previous section, \bea \tilde I_E &\equiv& \frac{2 \kappa^2 I_E}{4 \pi \Omega_2 \bar V_6 \bar
r_B^2}\nn &=& - 4 \frac{\beta}{4\pi \bar r_B}\left(\sqrt{\triangle_-
\triangle_+} - 1 \right) - \left(\frac{\bar r_+}{\bar
r_B}\right)^2,\nn &=& - 4 \bar b \left(\sqrt{(1 - x) \left(1 -
\frac{q^2}{x}\right)} - 1 \right) - x^2,\eea where in the last
equality we have used \eqn{reducedquantities} and \eqn{deltas}.
   Given the
  above form of $b_q (x)$, we have
  \be \label{bbehavior} b_q (x \rightarrow q) \rightarrow \infty, \quad b_q (x \rightarrow
  1) \rightarrow 0,\ee which is  the characteristic behavior of the
  corresponding $b_q (x)$ for black p-branes with $\tilde d > 2$. Our
  experience tells that the underlying phase structure contains the
  van der Waals-Maxwell liquid-gas type and there exists a critical
  charge $q_c$. When $q > q_c$, the equation of state $ \bar b = b_q
  (\bar x)$ has a unique solution $\bar x$ which gives a stable
  phase. When $q = q_c$, we have a second-order phase transition
  point, a critical point, at which we have a critical size $x_c$
  and a critical $b_c$ or critical temperature $T_c = 1/(4 \pi r_B
  b_c)$ and where there is no entropy change during the phase transition.
  When $q < q_c$, $b_q (x)$ has a maximum and a minimum, occurring
  at $x_{\rm max}$ and $x_{\rm min}$, respectively, with $0 < x_{\rm min} < x_{\rm max}
  <  1$. When the given $\bar b$ is between the minimum and the
  maximum, the equation of state $b_q (\bar x) = \bar b$  has
  three solutions $\bar x_1 < \bar x_2 < \bar x_3$ but only $\bar x_1$ and $\bar x_3$ give the locally stable phases, respectively,
  while $\bar x_2$ gives a unstable one since $b_q (x)$  has a
  positive slope at $\bar x = \bar x_2$. For each given $q < q_c$, there
  always exists a $b_t (q)$, which is a function of charge
  $q$ only, for which the corresponding phase at $\bar x_1$ and the one at
  $\bar x_3$ have the same free energy, therefore can coexist. The corresponding phase transition is
  a first-order one since it involves an
  entropy change during the phase transition (Note that the entropy
  is a function of $x$ and the transition involves a change of size
  $x$). Since $q < q_c$ is a one-parameter line segment, we
  therefore have a line of first-order phase transition terminating at
  the above mentioned second-order critical point. For given $q < q_c$, when $\bar b >
  b_t (q)$, the small size $\bar x_1$ phase has a lower free energy,
  therefore more stable while when $\bar b < b_t (q)$, on the other
  hand the large size $\bar x_3$ phase is more stable. The underlying
  phase is nothing but that of van der Waals-Maxwell liquid-gas type
  and in drawing the analogy, the small-size phase is like liquid
  one while the large size one like the gas phase. This behavior has
  been discussed in detail in \cite{Lu:2010xt} for charged black p-branes
  with $\tilde d > 2$ and it applies here, too.

  Let us go a bit further about the behavior of $b_q (x)$
  \eqn{reducedb}.  The $x_{\rm min}$ and $x_{\rm max}$ mentioned above are determined by
  $d b_q (x)/d x = 0 $, which gives \be \label{extremalcond}
  x^4 - \frac{2}{3} (1 + q^2) x^3 - 2 q^2 x^2 + 2 q^2 (1 + q^2) x -\frac{5}{3} q^4 = 0.\ee This is a quartic equation and has four roots
  in general. Since the sign of the last term (the constant term) in the equation is negative, therefore the number of negative roots is one or three.
  Note that $q < x < 1$ and if we use $f (x)$ to represent the left-hand side of the above equation, we have $f (q) = 4 q^2 (1 - q)^2/3 > 0$
  and $f (1) = (1 - q^2)^2/3 > 0$, both of which are positive. So if $ f (x) = 0$, i.e. \eqn{extremalcond}, has any solution in the
  region of $q < x < 1$, the number of them must be two.
  The function $f (x)$ has a unique minimum
  in the region of $q < x < 1$. We can show this by setting its first derivative
  vanish and this gives the equation \be x^3 - \frac{1 + q^2}{2} x^2 -
  q^2 x + \frac{1 + q^2}{2} q^2= 0,\ee which has three solutions $x = -
  q, q$ and $(1 + q^2)/2$, among which only the last solution falls
  in the region of interest. Further if we denote $f'' (x)$ as the second derivative of $f (x)$, we have
  \be f'' (\frac{1 + q^2}{2})
  = \frac{(1 - q^2)^2}{4} > 0,\ee therefore $f (x)$ has a unique
  minimum $f_{\rm min}$ occurring at $x = (1 + q^2)/2$, which implies that $f (x) = 0$ or \eqn{extremalcond} has either two roots when $f_{\rm min} < 0$ or none
  when $f_{\rm min} > 0$ in the region of $q < x < 1$. The value of $f_{\rm min}$ can be evaluated  as
  \be \label{fmin} f_{\rm min}  = - \frac{1}{48} (1 - q^2)^2\left[1
  + 2 \sqrt{5} q + q^2\right]\left(\sqrt{5} + 2 - q\right) \left(\sqrt{5} - 2 - q\right).\ee
  Given $0 < q < 1$, $f_{\rm min} > 0$ requires $q > \sqrt{5} - 2$ with which \eqn{extremalcond} has no solutions in the region of $q < x < 1$,
  therefore $b_q (x)$ decreases monotonically from the behavior of $b_q (x)$
  given in \eqn{bbehavior}. So the equation of state $\bar b = b_q (\bar x)$ has a unique
  solution $\bar x$ for each given $\bar b$ and the slope of $b_q (x)$ at $x = \bar x$ is negative, therefore giving a stable phase.
  $f_{\rm min} < 0$ requires $q < \sqrt{5} - 2$ and now \eqn{extremalcond} has two solutions $x_{\rm min} < x_{\rm max}$ in the region
  of $q < x < 1$ once again given the behavior of $b_q (x)$ \eqn{bbehavior}, the former giving the
  minimum of $b_q (x)$ and the latter the maximum. For each given $\bar b$ between the minimum and the maximum of $b_q (x)$, the equation of state
  $\bar b = b_q (\bar x)$ has three solutions $\bar x_1 < \bar x_2 < \bar x_3$ and exactly following the discussion given in \cite{Lu:2010xt}, we will have a van der
  Waals-Maxwell phase structure as described earlier and will not repeat it here.    $f_{\rm min} = 0$
  requires $q = q_c = \sqrt{5} - 2$ at which the two roots of \eqn{extremalcond} $x_{\rm
  min}$ and $x_{\rm max}$ coincide and so $x_{\rm min} = x_{\rm max} = x_c = (1 + q_c^2)/2 = 5 - 2
  \sqrt{5}$. Now the phase transition involves no entropy change and
  therefore it is a second-order critical point. At this point, not
  only we have the vanishing first derivative of $b_q (x)$,  its second derivative
  vanishes, too. The critical parameters can be collectively given
  as \be q_c = \sqrt{5} - 2, \quad x_c = 5 - 2\sqrt{5},\quad b_c =
  \frac{5^{5/4} (\sqrt{5} - 2)^{3/2}}{2},\ee where $b_c$ is obtained
  by substituting $x_c$ and $q_c$ into $b_q (x)$.  The above
  critical charge can also be obtained following \cite{Lu:2010xt}
  from the discriminant of \eqn{extremalcond} as
  \be \label{discriminant} \triangle = \frac{256}{27} q^6 (q^2 - 1)^4
  [(q - 2)^2 - 5][(q + 2)^2 - 5],\ee which vanishes only at $q_c =
  \sqrt{5} - 2$ (since $0 < q < 1$), implying the two positive roots being identical.
The characteristic behaviors, described above, of function $f (x)$
and $b_q (x)$ for $q > q_c, q = q_c$ and $q < q_c$ are given in Fig.
2  and Fig. 3, respectively. Note that we choose a smaller $q$ in
Fig. 3 than the one in Fig. 2 in the case of $q < q_c$ so that the
$b_q (x)$ curve shows clearly a minimum and a maximum.
\begin{figure}
\psfrag{a}{$x$} \psfrag{b}{$ f (x) $}
\begin{center}
\includegraphics{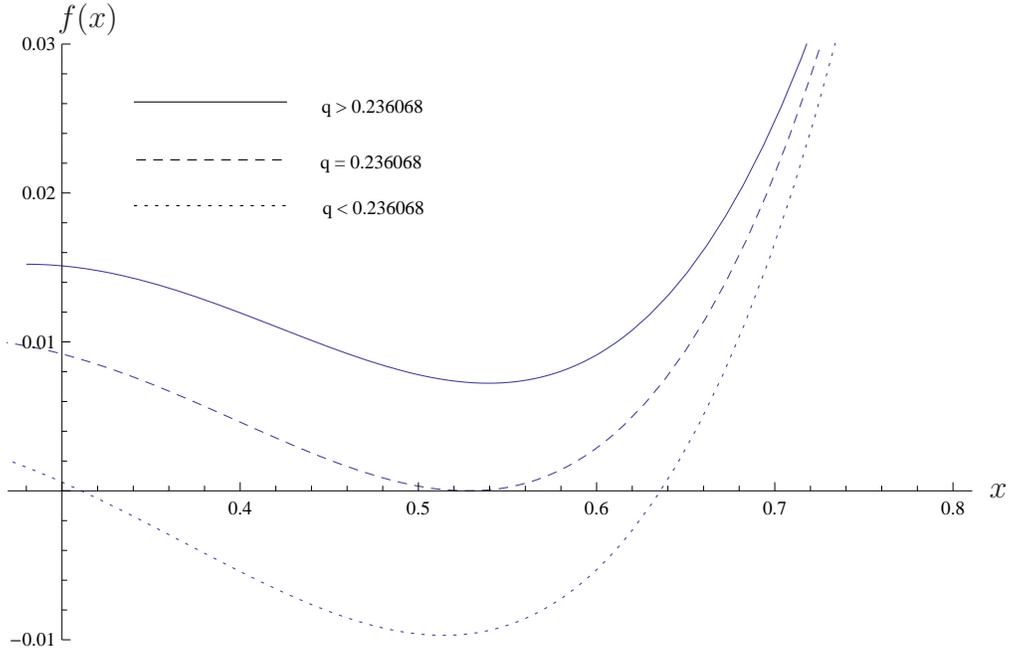}
\end{center}
  \caption{The characteristic behaviors of $ f(x) $ vs $x$ for $q > q_c, q = q_c$ and $q < q_c$ with $q_c = \sqrt{5} - 2 \approx 0.236068$
  as described in the text, respectively.}
\end{figure}

\begin{figure}
\psfrag{a}{$x$} \psfrag{b}{$b_q (x)$}
\begin{center}
\includegraphics{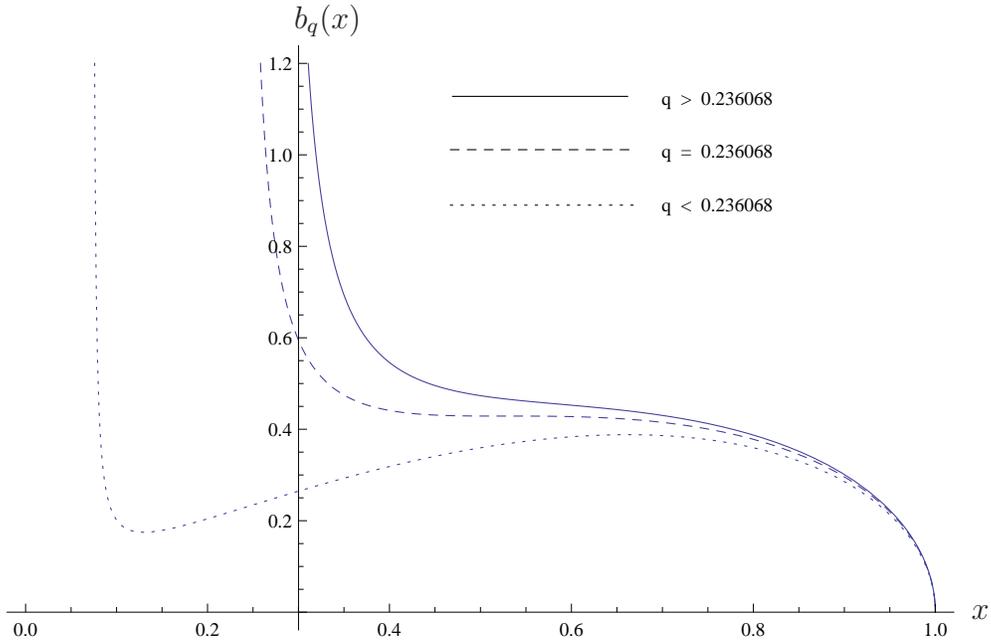}
\end{center}
  \caption{The characteristic behaviors of $b_q (x) $ vs $x$ for $q > q_c, q = q_c$ and $q < q_c$ with $q_c = \sqrt{5} - 2 \approx 0.236068$ as
  described in the text, respectively.}
\end{figure}
Due to the complication of the general configuration \eqn{bd0d6} for
charged black D0/D6 system, we use its two special cases, namely,
the $P = 0$ case and the $P = Q$ case to demonstrate: 1) the
delocalized D0 branes alone have the same phase structure as the
charged black D6 branes, and 2) adding the delocalized D0 branes to
the charged black D6 branes changes the phase structure of the
latter dramatically to the one, now containing a van der
Waals-Maxwell liquid-gas type structure, which does not occur when
either the delocalized D2 or D4 branes are added. It has long been
known (also obvious from the metric given in \eqn{bd0d6k}) that the
special $P = Q$ configuration, when dimensionally reduced along D6
worldvolume directions from $D = 10$ to $D = 4$, is nothing but the
$D = 4$ Reissner-Nordstr$\ddot{\rm o}$m one and its phase structure
in canonical ensemble has been analyzed in
\cite{Carlip:2003ne,Lundgren:2006kt}, though using a rather
different way from what we did above. It is interesting to find that
the inverse of local temperature $\beta (r_+)$ used there is same as
ours \eqn{case3beta} and so the phase structure is exactly the same
though the purpose here (showing how the added delocalized D0 branes
change qualitatively the phase structure of charged black D6 branes
dramatically) is completely different from theirs. Nevertheless this
reflects that the phase structure will remain the same for
dimensionally-reduced related systems, an expected result, in
addition to what we have found about T-dual related systems.

\section{Discussion and conclusion}
In this paper, we explore means which can be used potentially to
change or modulate qualitatively the phase structure of black
p-branes in canonical ensemble. Following \cite{Lu:2010xt}, we know
that the charged black branes with their $\tilde d = 2$ serve as a
borderline between the black branes with $\tilde d = 1$ and those
with $\tilde d > 2$ in phase structure. In a previous paper
\cite{Lu:2012rm}, we have studied the $\tilde d = 2$ case in $D =
10$ and the corresponding branes are charged black D5 branes.  The
delocalized charged black D-strings alone are found to have the
exact same phase structure as the charged black D5-branes without a
van-der Waals-Maxwell liquid-gas type structure. But when the two
are combined to form D1/D5 system, the resulting phase is much
richer and contains the liquid-gas type. We find that the physical
reason for this to occur is the existence of a non-vanishing
interaction between charged black D-strings and charged black
D5-branes, which differs from that between the delocalized black
D-strings or between black D5-branes. However, for the same type of
D2/D6 system, we don't find that adding the delocalized D2 branes to
the charged black D6 branes changes the phase structure of charged
black D6 branes that much and actually this only changes the phase
structure of latter to the one similar to that of charged black D5
branes, still containing no van der Waals-Maxwell liquid-gas type.

In general, adding delocalized D(p - 2) branes to Dp branes (see
footnote \ref{footnote1}) will not change the phase structure of the
latter at all. The only remaining possibility  for this purpose is
to consider adding the delocalized charged D0 branes to the charged
black D6 branes and it turns out that in this case, the phase
structure get modified to the required one.  Note that as discussed
in this paper the delocalized charged black D0 branes, D2 branes or
D4 branes each alone have the same phase structure as the charged
black D6 branes with $\tilde d = 1$ described in detail in
\cite{Lu:2010xt} (similar to the one for the chargeless case). As
discussed in the previous section, one possible reason for this is
that each of them have the same transverse dimensions $\tilde d + 2
= 3$ and the other more sounding reason is that they are all related
to each other by T-dualities (They all have the same isometries).
However, when each of them are combined with charged D6 branes to
form the corresponding charged black bound state D (6 - n)/D6 with
$n = 2, 4, 6$, the resulting phase structures are rather different,
a bit of surprise. When $n = 2$, the phase structure of the charged
black D4/D6 remains the same as that of individual constituents, the
naively expected one. When $n = 4$, the resulting phase structure of
the charged black D2/D6 is changed to the one similar to that of
charged black D5 branes with $\tilde d = 2$ given in
\cite{Lu:2010xt, Lu:2012rm}, but in essence the phase structure is
still in the same class as that for the chargeless case as described
earlier, containing no van der Waals-Maxwell liquid-gas type. Only
for $n = 6$, the resulting phase structure got changed dramatically
to the one similar to that of charged black p-branes with $\tilde d
> 2$, containing now a van der Waals-Maxwell liquid-gas type as
described in the previous section. We all know (for example, see
\cite{polchinski-v2}) that when D(6 - n) and D6 are combined to form
charged black bound state D(6 - n)/D6, the interaction between
constituent D(6 - n) and D6 are different for different $n = 2, 4,
6$, respectively. For example, adding the delocalized charged D4 to
D6 gives rise to an additional attractive interaction while adding
delocalized charged D0 to D6 gives rise to an additional repulsive
interaction.  As mentioned earlier, the repulsive interaction due to
the added D6 brane charge is not sufficient to change its existing
phase structure and additional repulsive interaction is needed for
this purpose. Therefore it is not surprised for us to have the above
phase structure for the corresponding system at least qualitatively.
So the above different phase structure for each case should be
expected qualitatively to be due to the different interaction
between the constituent branes.  It would be interesting to explore
the exact or quantitative correlation between the interaction and
the phase structure and we hope to return this issue elsewhere. We
also find that the black systems related by T-duality (to be
precise, that two black systems are related by T-duality here means
that the corresponding two gravity configurations can be related to
each other by T-duality in the sense described earlier and they must
have the same isometries. For example, the delocalized charged black
D0 configuration \eqn{bd0} is T-dual to the charged black D6
configuration \eqn{bd6}.) or dimensionally-reduced black related
systems have the same phase structure. Though expected, it is nice
to confirm this.

\section*{Acknowledgements:}

 We would like to thank Shibaji Roy for help in polishing the revision. We acknowledge
support by grants from the Chinese Academy of Sciences and grants
from the NSF of China with Grant No : 10975129 and 11235010.


\begin{thebibliography}{99}

\bibitem{Lu:2012rm}
  J.~X.~Lu, R.~Wei and J.~Xu,
  ``The phase structure of black D1/D5 (F/NS5) system in canonical ensemble,'' JHEP {\bf 1212},
  012(2012). arXiv:1210.0708 [hep-th].


\bibitem{Carlip:2003ne}
  S.~Carlip and S.~Vaidya,
  ``Phase transitions and critical behavior for charged black holes,''
  Class.\ Quant.\ Grav.\  {\bf 20} (2003) 3827
  [arXiv:gr-qc/0306054].

\bibitem{Lundgren:2006kt}
  A.~P.~Lundgren,
  ``Charged black hole in a canonical ensemble,''
  Phys.\ Rev.\  D {\bf 77}, 044014 (2008)
  [arXiv:gr-qc/0612119].

\bibitem{Hawking:1982dh}
  S.~W.~Hawking and D.~N.~Page,
  ``Thermodynamics Of Black Holes In Anti-De Sitter Space,''
  Commun.\ Math.\ Phys.\  {\bf 87}, 577 (1983).

\bibitem{Chamblin:1999tk}
  A.~Chamblin, R.~Emparan, C.~V.~Johnson and R.~C.~Myers,
  ``Charged AdS black holes and catastrophic holography,''
  Phys.\ Rev.\  D {\bf 60}, 064018 (1999)
  [arXiv:hep-th/9902170].

 \bibitem{Chamblin:1999hg}
  A.~Chamblin, R.~Emparan, C.~V.~Johnson and R.~C.~Myers,
  ``Holography, thermodynamics and fluctuations of charged AdS black holes,''
  Phys.\ Rev.\  D {\bf 60}, 104026 (1999)
  [arXiv:hep-th/9904197].

\bibitem{York:1986it}
  J.~W.~York,
  ``Black hole thermodynamics and the Euclidean Einstein action,''
  Phys.\ Rev.\  D {\bf 33}, 2092 (1986).


\bibitem{Braden:1990hw}
  H.~W.~Braden, J.~D.~Brown, B.~F.~Whiting and J.~W.~.~York,
  ``Charged black hole in a grand canonical ensemble,''
  Phys.\ Rev.\  D {\bf 42}, 3376 (1990).


\bibitem{Witten:1998zw}
  E.~Witten,
  ``Anti-de Sitter space, thermal phase transition, and confinement in gauge theories,''
  Adv.\ Theor.\ Math.\ Phys.\  {\bf 2}, 505 (1998)
  [arXiv:hep-th/9803131].


\bibitem{Lu:2010xt}
  J.~X.~Lu, S.~Roy, Z.~Xiao,
  ``Phase transitions and critical behavior of black branes
in canonical ensemble,''
  JHEP {\bf 1101}, 133 (2011).
  [arXiv:1010.2068 [hep-th]].

 \bibitem{Lu:2010au}
   J.~X.~Lu, S.~Roy, Z.~Xiao, ``Phase structure of black branes in grand canonical ensemble,''
JHEP {\bf 1105}, 091(2011) [arXiv:1011.5198 [hep-th]].

\bibitem{Cai:2000hn}
  R.~-G.~Cai and N.~Ohta,
  ``Noncommutative and ordinary superYang-Mills on (D(p - 2), D p) bound states,''  JHEP {\bf 0003}, 009 (2000)  [hep-th/0001213].

\bibitem{Taylor:1997ay}
  W.~Taylor,
  ``Adhering zero-branes to six-branes and eight-branes,''  Nucl.\ Phys.\ B {\bf 508}, 122 (1997)  [hep-th/9705116].

\bibitem{Brandhuber:1997tt}
  A.~Brandhuber, N.~Itzhaki, J.~Sonnenschein and S.~Yankielowicz,
  ``More on probing branes with branes,''  Phys.\ Lett.\ B {\bf 423}, 238 (1998)  [hep-th/9711010].

\bibitem{Dhar:1998ip}
  A.~Dhar and G.~Mandal,
  ``Probing four-dimensional nonsupersymmetric black holes carrying D0-brane and D6-brane charges,''
   Nucl.\ Phys.\ B {\bf 531}, 256 (1998)  [hep-th/9803004].

\bibitem{Gibbons:1975kk}
  G.~W.~Gibbons,
  ``Vacuum Polarization and the Spontaneous Loss of Charge by Black Holes,''  Commun.\ Math.\ Phys.\  {\bf 44}, 245 (1975).

\bibitem{Gibbons:1985ac}
  G.~W.~Gibbons and D.~L.~Wiltshire,
  ``Black Holes in Kaluza-Klein Theory,''  Annals Phys.\  {\bf 167}, 201 (1986)  [Erratum-ibid.\  {\bf 176}, 393 (1987)].


\bibitem{Lu:2011da}
J.~ X.~Lu, S. Roy, Z.~Xiao,
  ``The enriched phase structure of black branes in canonical
  ensemble,"
  Nucl.\ Phys.\ {\bf B854}, 913(2012).



\bibitem{Horowitz:1991cd}
  G.~T.~Horowitz and A.~Strominger,
  ``Black strings and P-branes,''  Nucl.\ Phys.\ B {\bf 360}, 197 (1991).


\bibitem{Duff:1993ye}
  M.~J.~Duff and J.~X.~Lu,
  ``Black and super p-branes in diverse dimensions,''  Nucl.\ Phys.\ B {\bf 416}, 301 (1994)  [hep-th/9306052].

\bibitem{Duff:1994an}
  M.~J.~Duff, R.~R.~Khuri and J.~X.~Lu,
``String solitons,''  Phys.\ Rept.\  {\bf 259}, 213 (1995)  [hep-th/9412184].

\bibitem{Sheinblatt:1997nt}
  H.~J.~Sheinblatt,
  ``Statistical entropy of an extremal black hole with 0-brane and 6-brane charge,''  Phys.\ Rev.\ D {\bf 57}, 2421 (1998)  [hep-th/9705054].

\bibitem{Gibbons:1976ue}
G.~W.~Gibbons and S.~W.~Hawking, ``Action Integrals And Partition
Functions In Quantum Gravity,'' Phys.\ Rev.\  D {\bf 15}, 2752
(1977).

\bibitem{Bergshoeff:1995as}
  E.~Bergshoeff, C.~M.~Hull and T.~Ortin,
  ``Duality in the type II superstring effective action,''  Nucl.\ Phys.\ B {\bf 451}, 547 (1995)  [hep-th/9504081].



\bibitem{Breckenridge:1996tt}
  J.~C.~Breckenridge, G.~Michaud and R.~C.~Myers,
  ``More D-brane bound states,''  Phys.\ Rev.\ D {\bf 55}, 6438 (1997)  [hep-th/9611174].



\bibitem{Lu:2005ju}
  J.~X.~Lu and S.~Roy,
  ``Non-SUSY p-branes delocalized in two directions, tachyon condensation and T-duality,''  JHEP {\bf 0506}, 026 (2005)
  [hep-th/0503007].

\bibitem{polchinski-v2}
J.~ Polchinski, ``Superstring Theory¡±, Vol. 2, Cambridge: Cambridge
University Press (1998)

















\end{thebibliography}
\end{document}